\documentclass[a4paper,twoside,twocolumn]{article}
\pdfoutput=1

\newif\iftr 
\trtrue     

\usepackage{ifpdf}  
\ifpdf
    \usepackage[pdftex]{graphicx}
    \usepackage[colorlinks,linkcolor=blue,citecolor=blue,urlcolor=blue]{hyperref}
    \DeclareGraphicsExtensions{.pdf}
\else
    \usepackage{graphicx}
    \usepackage[hypertex]{hyperref}  
\fi

\usepackage[disable]{todonotes} 
\newcommand{\bob}[1]{\todo[color=olive!40,inline]{Bob: #1}}

\usepackage{
url       %
,fancyhdr 
,lastpage 
,enumerate 
,amsmath  
,amsfonts 
,amssymb  
, enumitem 
, xfrac    
}
\usepackage[hang]{footmisc} 
\makeatletter
\def\namedlabel#1#2{\begingroup
	#2%
	\def\@currentlabel{#2}%
	\phantomsection\label{#1}\endgroup
}
\makeatother
\usepackage{balance}

\usepackage[noindent, arraystretch, fullpage]{setlengths} 
\usepackage{
own         
}

\usepackage{amsthm}

\newtheorem*{test*}{Test}

\graphicspath{{images/}}

\newcommand*{\metaauthori}{Bob Briscoe}
\newcommand*{\metashorttitle}{PI\(^2\) Parameters}
\newcommand*{\metatitle}{PI\(^2\) Parameters}
\newcommand*{\metano}{TR-BB-2021-001}
\newcommand*{\metakeywords}{Data Communication, Networks, Internet, Performance, Latency, Congestion Control, Congestion Avoidance, QoS, Ultilization, Scaling, Algorithms, Implementation, Design, AQM, Standards}

\newcommand*{\metamaili}{\href{mailto:research@bobbriscoe.net}{research@bobbriscoe.net}}
\newcommand*{\metaaddress}{}

\newcommand*{\metaversion}{03}
\newcommand*{\metadate}{27 Oct 2021}

\hypersetup{                       
     pdfauthor = {\metaauthori
     },
     pdftitle = {\metashorttitle},
     pdfsubject = {},
     pdfkeywords = {\metakeywords}
}%

\title{\metatitle}%
\author{\metaauthori%
\thanks{\metamaili, %
\metaaddress}%
}
\date{\metadate}%

\fancypagestyle{first}{%
\fancyhead[LO,RE]{}%
\fancyhead[LE,RO]{}%
\fancyhead[C]{}%
}%

\begin{document}
\bibliographystyle{alpha}%


\maketitle%
\thispagestyle{first}

\begin{abstract}
{\small\noindent%

This report gives the reasoning for the setting of the target queue
delay parameter in the reference
Linux implementation of the PI\(^2\) AQM.
}      
\end{abstract}
\section{Introduction}\label{pi2param_intro}

This report explains the reasoning behind the setting of the queue delay \texttt{target}
in the reference Linux implementation of the PI\(^2\)
AQM\footnote{\url{https://github.com/L4STeam/sch_dualpi2_upstream}}. This
setting is documented as a pseudocode example in Figure 2 (in Appendix A) of the IETF
specification of the Coupled DualQ AQM~\cite{Briscoe15e:DualQ-Coupled-AQM_ID}.
In both cases, the PI\(^2\) AQM is used for the Classic queue within the
dual-queue structure called DualPI2. Nonetheless, the parameter settings for PI\(^2\)
discussed here apply irrespective of whether a PI\(^2\) AQM stands alone or
within a dual-queue structure. The discussion of the \texttt{target} parameter
also applies to a PIE AQM~\cite{Pan_PIE_2013}.

Similar reasoning for the parameter settings was behind the technical report
produced in 2015~\cite{DeSchepper15b:DCttH_TR} to support standardization of the
Coupled DualQ AQM. The present report spells
out all the details that were glossed over at that time, and adds some more recent
analysis, resulting in a slightly higher figure. 

The task for this report is to choose a compromise default for
\texttt{target} that minimizes queue delay for Classic traffic without causing
under-utilization over path RTTs that are commonly experienced by most Internet
users.

\section{Terminology}\label{pi2param_terms}

\begin{figure}
	\centering
	\includegraphics[width=0.8\columnwidth]{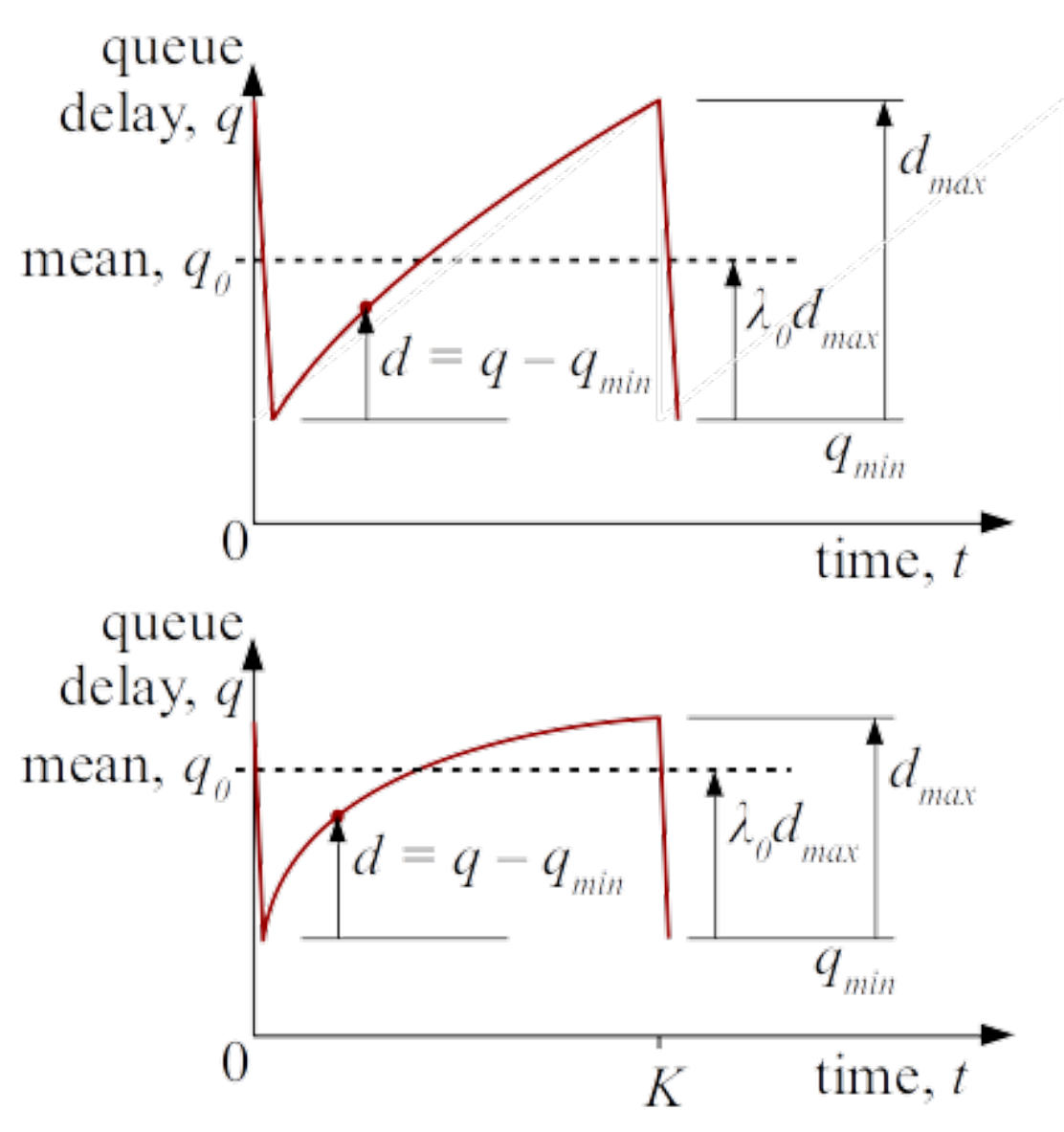}
	\caption{Definition of terms}\label{fig:pi2param-terms}
\end{figure}

The schematic plots of one cycle of queue delay against time for two different
congestion controllers (Reno and Cubic) in \autoref{fig:pi2param-terms} define
our terminology. We use \(q(t)\) as the time-varying queue
delay in units of time, and \(d(t)\) as the additional queue delay above the
minimum, that is \(d(t) = q(t)-q_\mathrm{min}\). \(q_0\) is the mean queue
delay for a particular geometry of sawtooth and \(d_\mathrm{max}\) is the amplitude of the
cycles, in units of time. The fraction, \(\lambda_0\)
of the amplitude that sits below the average depends solely on the geometry of
the sawtooth curve. A related fraction, \(\lambda\) (not shown) is defined as the fraction of
the amplitude that sits below the AQM's operating point, \texttt{target}.

The instantaneous RTT, \(R(t)\), varies because it consists of the constant base
delay of the path, \(R_b\), and variable queue delay \(q(t)\), that is \(R(t) =
R_b+q(t)\). Alternatively, \(R(t)=R_\mathrm{min}+d(t)\).

A Classic congestion controller's min window is related to its max window by the
multiplicative factor, \(b\), of the congestion controller: \(W_\mathrm{min} = b
W_\mathrm{max}\). This leads the RTT to cycle between \(R_\mathrm{min}\) and
\(R_\mathrm{max}\), around the average \(R_0\). Similarly queue delay cycles
around \(q_0\) between \(q_\mathrm{min}\) and \(q_\mathrm{max}\).

The IETF's specification of Cubic~\cite{Rhee18:Cubic_RFC} uses \(\beta\) for the
multiplicative decrease factor, but we use \(b\) to avoid confusion with the
proportional gain factor \(\beta\) of a PI AQM. Similarly, we use \(a\) (rather
than \(\alpha\)) for the additive increase factor of Reno, or of Cubic in its
Reno-friendly mode. Where necessary, we use the subscripts 'r' or 'c' to
distinguish parameters used by Reno or Cubic in Reno mode (`CReno').

\section{Scaling of Queue Variation}\label{pi2param_Scaling}

\begin{description}[nosep]
	\item[\namedlabel{itm:pi2param_one_long_flow}{Assumption 1}:] We are interested
	in the operating point that the queue cycles around under
	stable conditions, so we consider only long-running flows and fixed capacity
	links. Within this assumption of a stable environment, we consider a single flow as
	the worst-case for queue variability (and a fairly common case in access link
	bottlenecks). Given the long-running flow assumption, we also assume packet size is 1500\,B, where a feel for a packet rate is given as a bit rate.
\end{description}

The schematics in \autoref{fig:pi2param-scaling} show how different Linux
congestion controls vary the queue delay of a single flow around the mean and
how the variation scales with base RTT and link capacity. The scales of the
plots are all the same, but actual numerical values of queue delay and time are
irrelevant for this visualization.

The following two subsections consider how queue variation due to a single flow
scales with base RTT and with link capacity. Then
\autoref{pi2param_Sawtooth_Geometry} discusses the geometry, position and prevalence of
different sawtooth shapes.

\subsection{Scaling of Q Variation with RTT}\label{pi2param_Scaling_RTT}
\begin{figure}[t]
	\centering
	\includegraphics[width=\columnwidth]{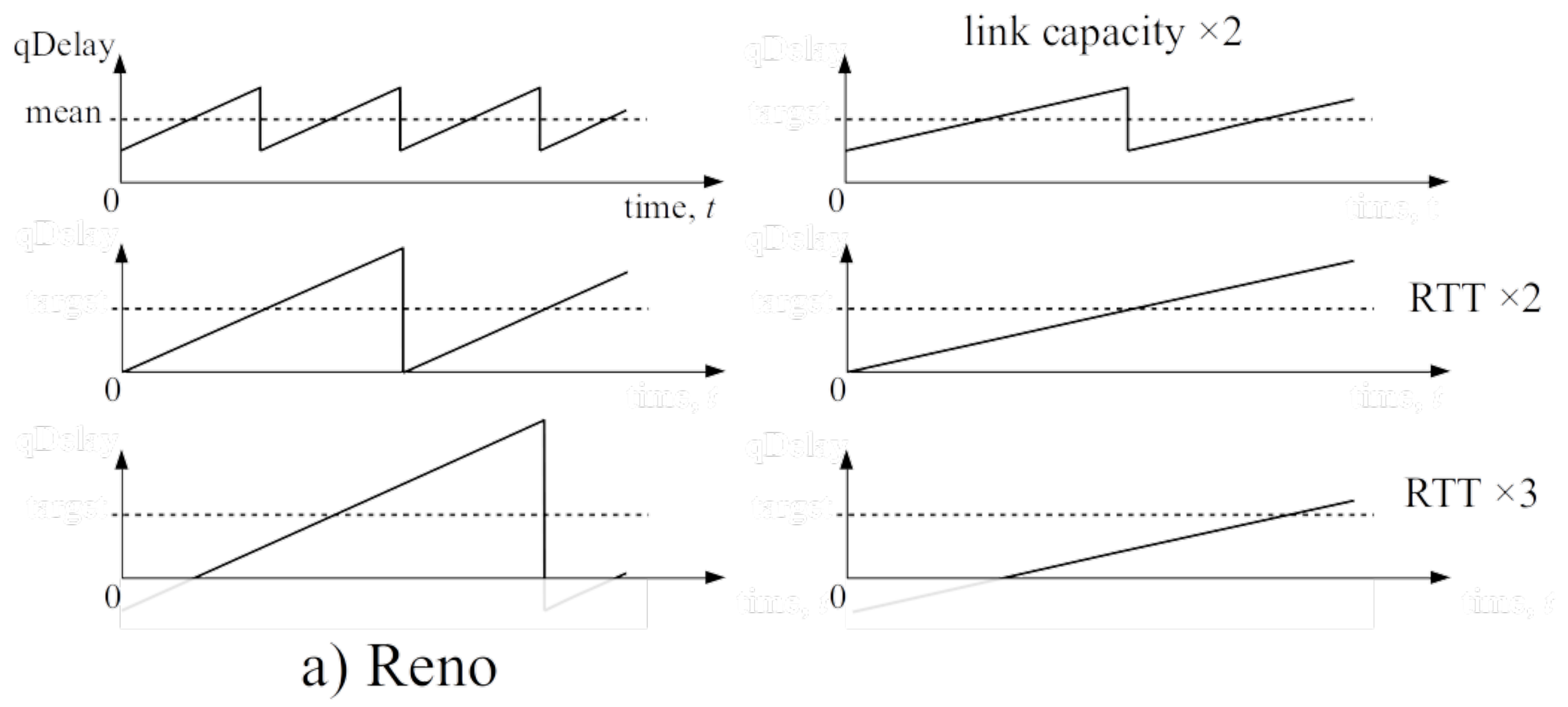}
	\includegraphics[width=\columnwidth]{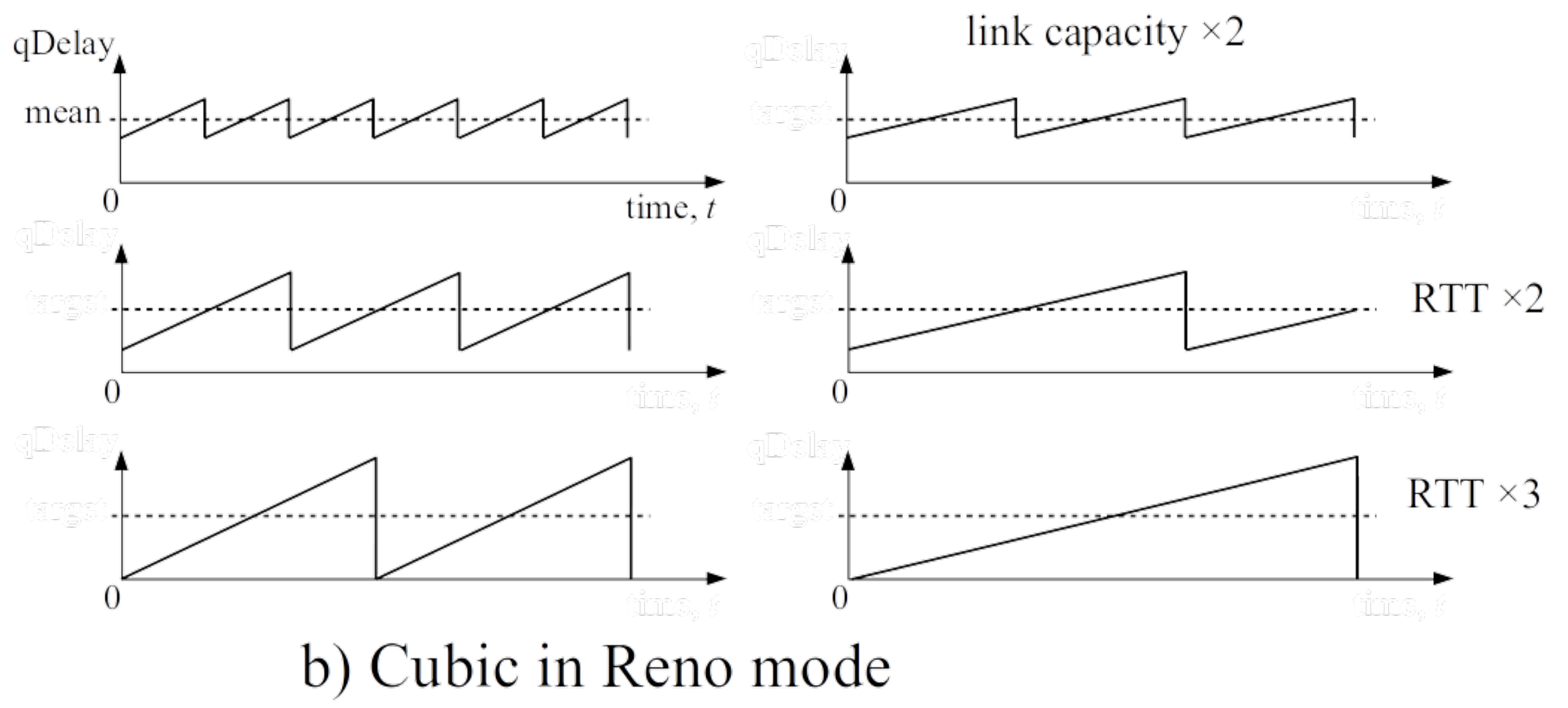}
	\includegraphics[width=\columnwidth]{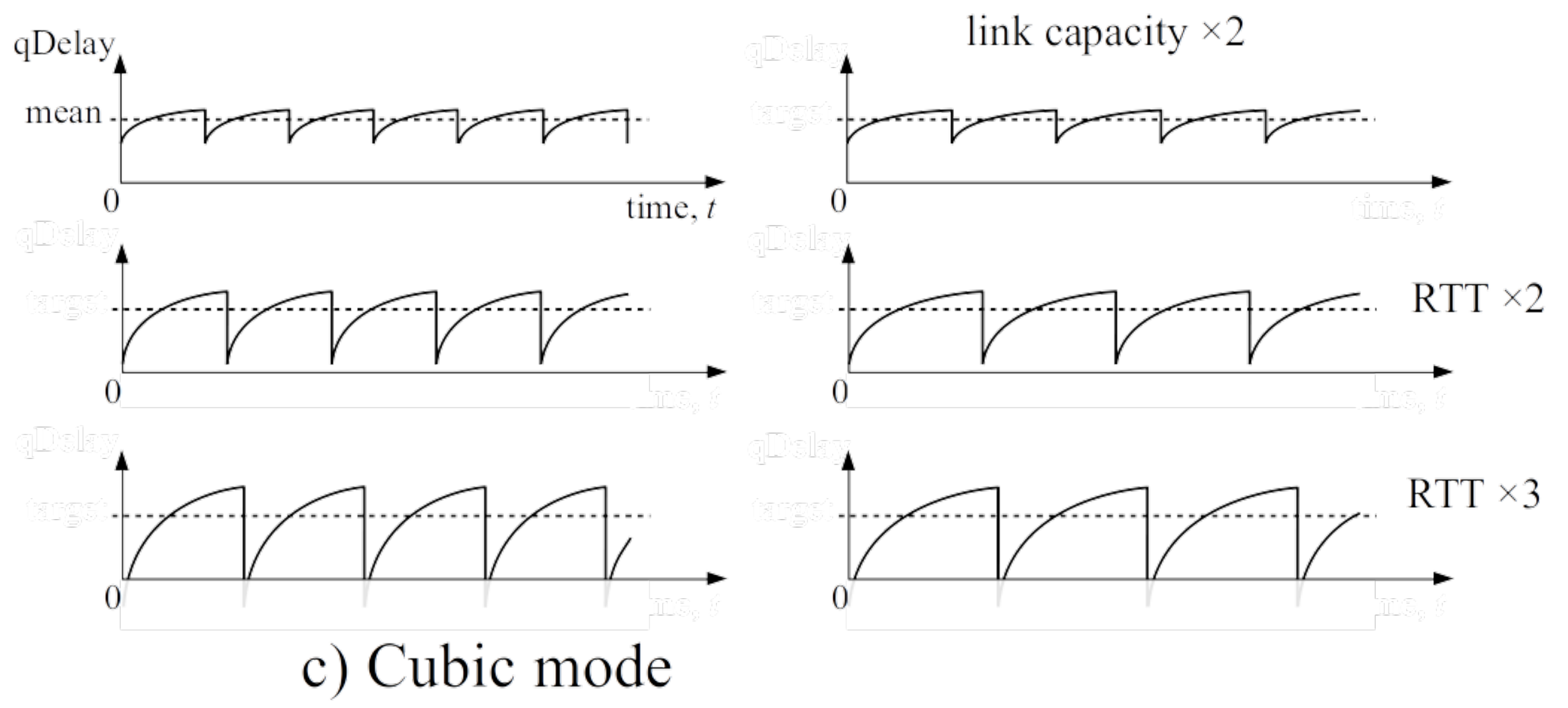}
	\caption{Scaling of queue delay variability with average RTT (increasing
	downwards) and link capacity (increasing to the
	right)}\label{fig:pi2param-scaling}
\bob{Update figure for corrected CReno}
\end{figure}

\begin{description}[nosep]
	\item[\namedlabel{itm:pi2param_busy}{Assumption 2}:] Our goal is to find a
	value of \texttt{target} that prevents the queue from completely draining at
	the bottom of each sawtooth cycle. Therefore, our analysis
	assumes that is the case because, when it is not, the analysis does not need to
	apply. So, we can assume constant delivered packet rate, \(r\), given we also 
	assume constant link capacity for simplicity
	(despite being unrealistic). Then, given that the window, \(W = r*R\), the RTT,
	\(R\), varies in direct proportion to the window.

	\item[\namedlabel{itm:pi2param_instant}{Assumption 3}:] As a first-order
	approximation, we assume that queue delay tracks the window instantly, even
	though it actually takes a round trip to catch up. And we don't consider
	smoothing of window reductions, e.g.\ Proportion Rate Reduction. These
	approximations overestimate the amplitude of queue variation a little,
	especially when the window reduces sharply then increases sharply (as it does
	when Cubic responds to congestion). However, these approximations are close
	enough for our purposes.
\end{description}

By \ref{itm:pi2param_busy},
\begin{align}
R_\mathrm{min}	&= b R_\mathrm{max}\label{eqn:Rmin_max}
\end{align}
Then the algebra below shows that sawtooth amplitude is related to average RTT by a
constant factor,
\begin{align}
			R_0	&= R_\mathrm{min} + \lambda_0(R_\mathrm{max}-R_\mathrm{min})\notag\\
				&= R_\mathrm{max}(\lambda_0+b-\lambda_0  b)\label{eqn:R0_max}\\
d_\mathrm{max}	&= R_\mathrm{max} - R_\mathrm{min}\notag\\
				&= (1-b)R_\mathrm{max}\notag\\
				&= \frac{(1-b)}{(\lambda_0+b-\lambda_0  b)}R_0.\notag
\end{align}
This is why, starting at the top left and working down the schematics for each
congestion control in \autoref{fig:pi2param-scaling}, it is shown that the
amplitude of the queue variation grows linearly with average RTT.\footnote{At
least, it does while the sawteeth do not drain the queue completely, by
\ref{itm:pi2param_busy}. Where this assumption breaks down---in the plots
labelled `RTT \(\times3\)' for a) Reno and c) Cubic mode---for visualization
purposes light grey traces extrapolate where the plots would be if the queue
could be negative.} This linear scaling of queue variability with RTT only
relies on multiplicative decrease, so it is just as true for either mode of
Cubic as it is for Reno.
\begin{figure*}
	\centering
	\includegraphics[width=0.32\linewidth]{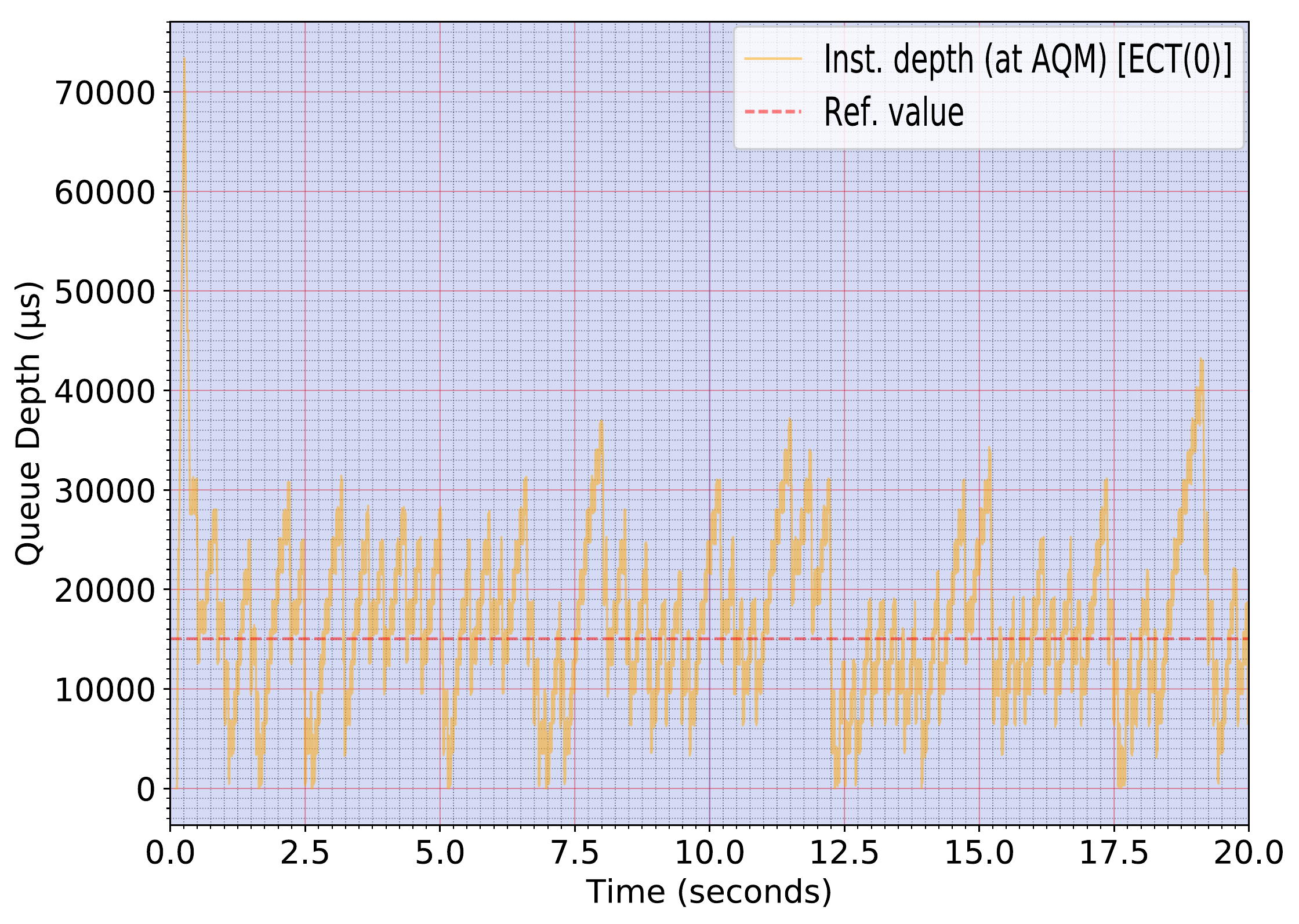}
	\includegraphics[width=0.32\linewidth]{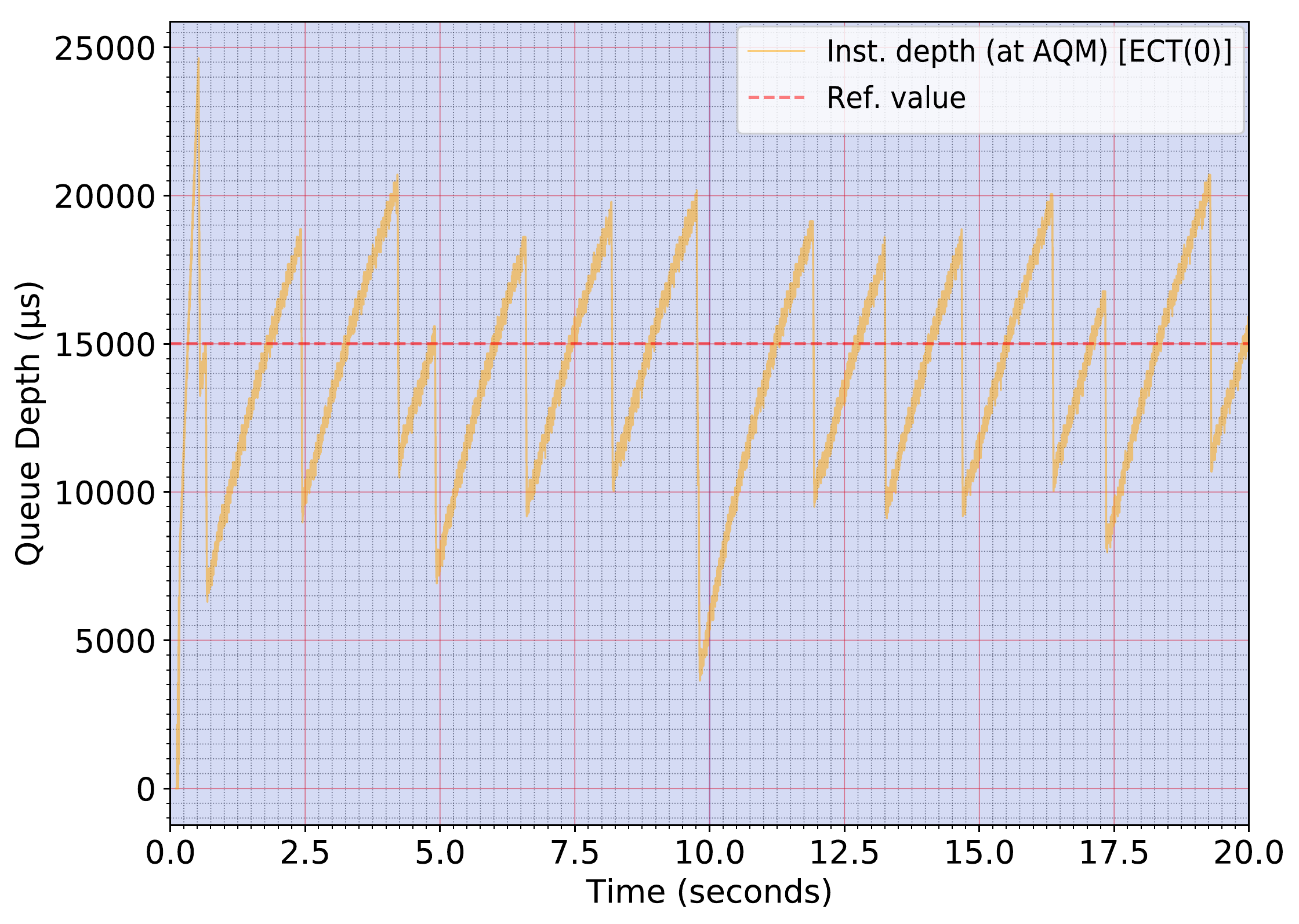}
	\includegraphics[width=0.32\linewidth]{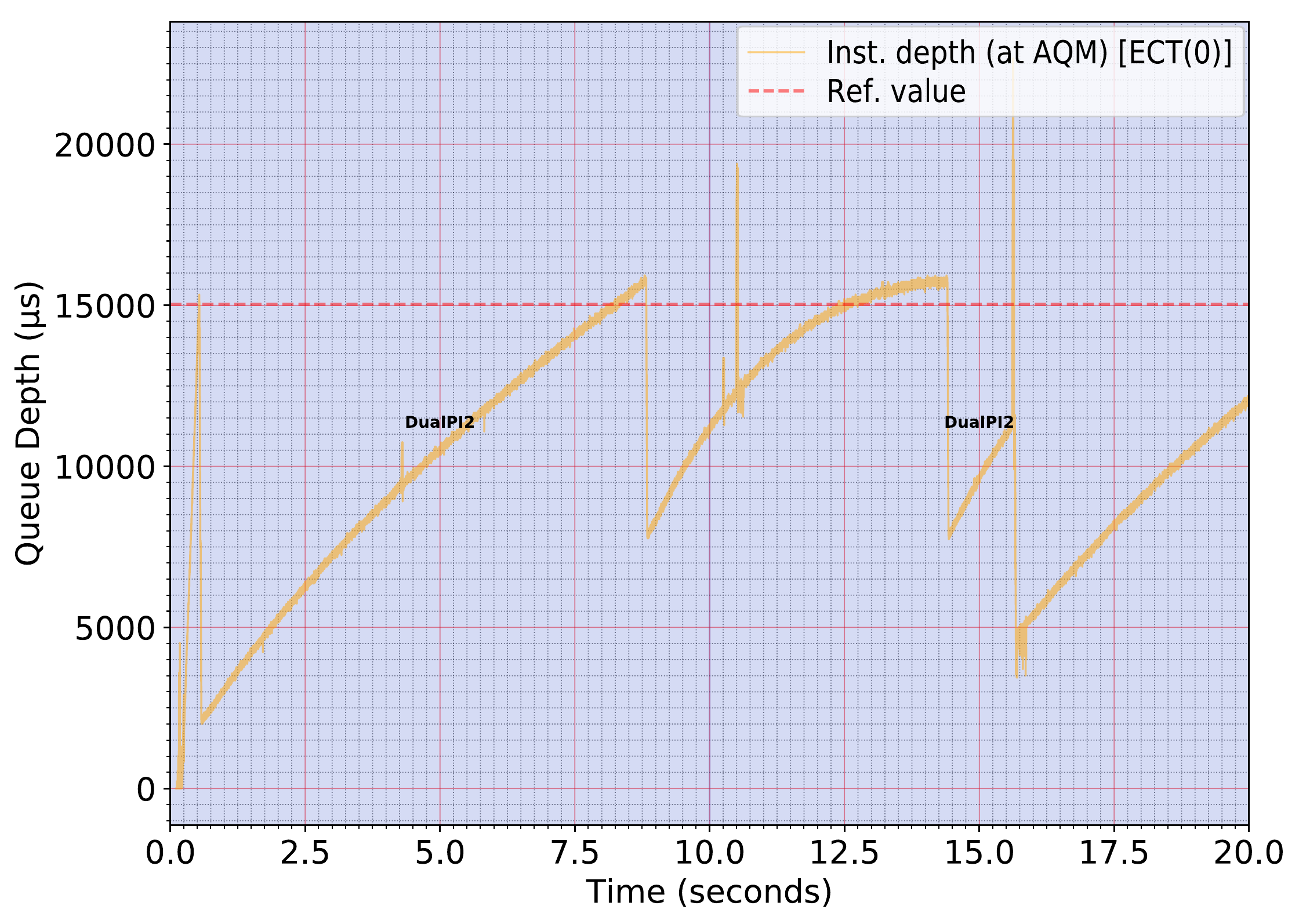}
	\caption{Transition of the position of a Cubic/CReno sawtooth relative to the
		PI\(^2\) AQM \texttt{target} (15\,ms). Base RTT: 10\,ms, Link rate (left to right): 4\,Mb/s, 40\,Mb/s, 200\,Mb/s.}\label{fig:pi2param-TcTransition-expts}
\end{figure*}
\begin{figure*}
	\centering
	\includegraphics[width=0.7\linewidth]{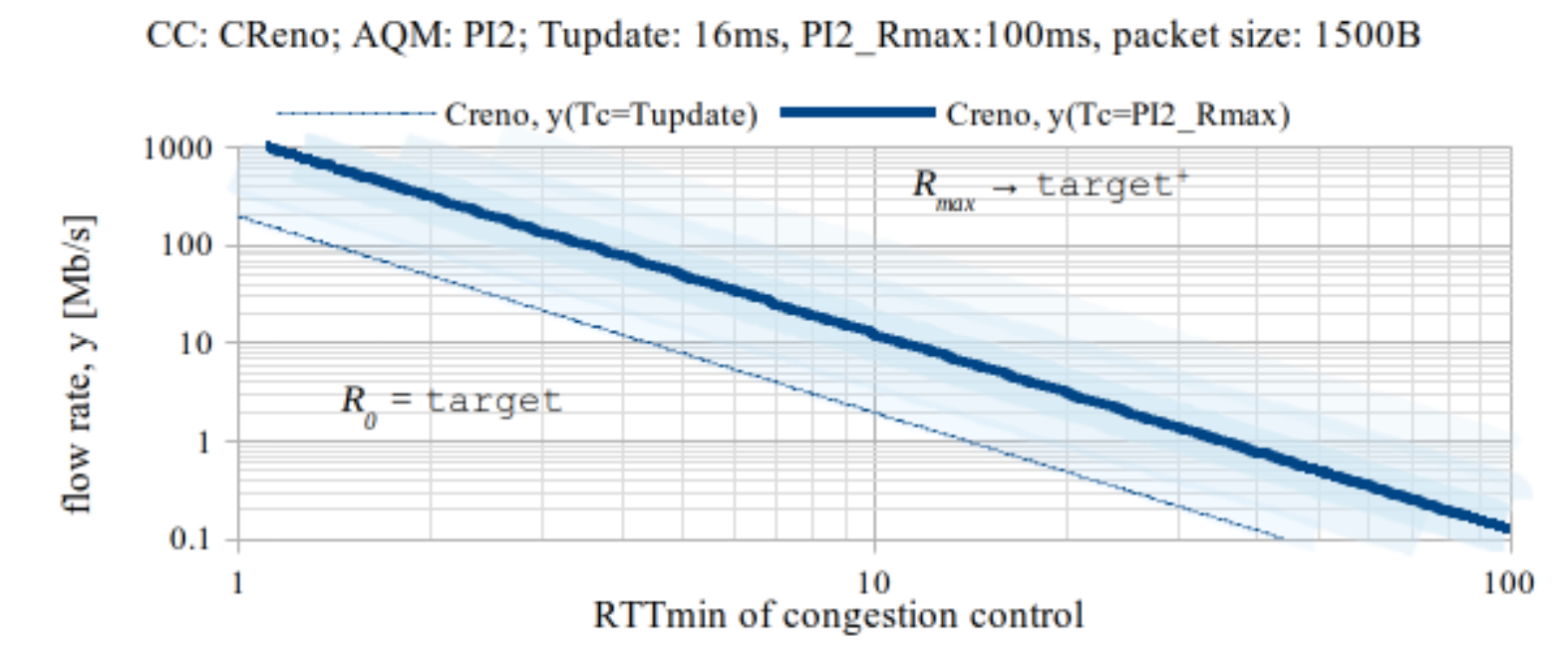}
	\caption{Transition (shaded) of relationship between CReno sawteeth and
		PI\(^2\) AQM \texttt{target}. Below the range, CReno sawteeth average at
		\texttt{target}. Above the range they peak at
		\texttt{target}}.\label{fig:pi2param-TcTransition}
\end{figure*}

\subsection{Scaling of Queue Variation with Link Capacity}\label{pi2param_Scaling_Capacity}

More link capacity allows either more flows or more throughput per flow. But in
the edge links giving access to the Internet, which tend to be the bottleneck
links, the number of simultaneous flows is still low, and single lone flows
remain common~\cite{Rajiullah15:WhatUseTopSpeed_TR}.

If link capacity doubles, the delivered packet rate (and the average window) of
a single flow doubles too. Nonetheless, the PI\(^2\) AQM holds the average RTT,
\(R_0\), at the same operating point\footnote{\label{note:rtt-const}This
	assertion will be qualified in \S\,\ref{pi2param_Sawtooth_Position}}.
Then, for a particular congestion control, by \autoref{eqn:Rmin_max} \&
\autoref{eqn:R0_max} the max and min RTT are related to the
average RTT by constant factors, so they also remain unchanged.

Therefore, \(q_\mathrm{max}\) and \(q_\mathrm{min}\) also remain unchanged as
link capacity scales (shown in the right-hand column of
\autoref{fig:pi2param-scaling}).\footnote{See footnote \ref{note:rtt-const}}
This also means that queue delay variation when the upstream is filled by a
single flow is no different from the variation when a single flow fills the
downstream, even if the capacity is asymmetric.

Incidentally, the scaling of the cycle duration (along the horizontal time axis
in \autoref{fig:pi2param-scaling}) is not directly relevant to queue variation,
but  \autoref{pi2param_Scaling_Cycle} briefly explains why it is indirectly relevant in two ways.

\subsection{Sawtooth Position}\label{pi2param_Sawtooth_Position}

As link rate scales, the assertion that the PI\(^2\) AQM holds the average RTT,
\(R_0\), at \texttt{target} needs qualification. It is true when the sawtooth cycle is short, as shown on the left of \autoref{fig:pi2param-TcTransition-expts}. However, as the duration of the cycles increases from left to right, the sawteeth are increasingly pushed down so that only their tips touch the target.

PI\(^2\) samples the queue and
updates its drop probability every update time (\texttt{Tupdate} = 16\,ms by
default). The PI controller determines the size of each of its probability alterations to bring delay under control at two different timescales that are controlled by the gain factors \(\alpha\) and \(\beta\): the \textbf{I}ntegral term of the controller brings the standing queue back to \texttt{target} within about \(\mathrm{PI2}_\mathrm{Rmax}\) (default 100\,ms); while the \textbf{P}roportional term catches variations an order of magnitude faster. 

At low bandwidth-delay product (BDP)\footnote{Strictly, for AIMD congestion controls, the effect depends on the product of packet rate and the square of delay, so we should say at low BDDP.}, when the duration of each sawtooth (the `recovery time') is less than the AQM's update time, \texttt{Tupdate}, the AQM will never be able to track the rise and fall of each sawtooth; so its probability will remain steady and it will hold only the average level of delay at \texttt{target}. For CReno, this is the region under the fine dashed line that delineates the floor of the shaded transition range, which is plotted in \autoref{fig:pi2param-TcTransition} over a space 
covering a range of flow rate-RTT combinations.

When the recovery time is well above the AQM's maximum design RTT (\(PI2_\mathrm{Rmax}\)), the AQM will be able to track the sawteeth fairly closely. So when queue delay reaches target, the AQM will emit a drop or ECN-mark and the subsequent sawteeth will settle with their peaks just above \texttt{target}. For CReno, this is the region well above the shaded transition range shown in \autoref{fig:pi2param-TcTransition}.

When the recovery time of the sawteeth is the same as the time that the AQM takes to converge on its \texttt{target} (\(\mathrm{PI2}_\mathrm{Rmax}\), default 100\,ms), the AQM can start to track the variations in the sawtooth, but not quickly enough to keep up. For CReno, this is shown as the thick dark blue curve in the middle of the shaded transition range in \autoref{fig:pi2param-TcTransition}. In the shaded transition region around this central curve the sawtooth settles with \texttt{target} somewhere between its average and its peak. 

This effect is less pronounced with Cubic sawteeth than AIMD, because Cubic
sawteeth spend more of their duration close to the average, with only a brief
large deviation at the start. However, for sufficiently long sawteeth the AQM
will still track the sawtooth itself, not just the average.

It is hard to derive
the position of the sawteeth analytically, so we resort to estimating the fraction of the
sawtooth amplitude empirically (visually) from large numbers of time series plots. This, in turn, is hard given the point at which the sawtooth reduces is
randomized, by design. Nonetheless, 
once the cycle time is well above the transition region,
on average AIMD sawteeth tend to settle with
the AQM target about 90\% of the amplitude above the minimum. Whereas Cubic
sawteeth tend to settle lower down the sawtooth---nearer to 85\% of the
amplitude (the average height of a cubic cycle is 75\% of its amplitude according to \autoref{eqn:lambda_cubic} in \autoref{pi2param_cubic_R_avg}).

If the BDP of a Cubic flow is high enough to put it into true Cubic mode, its long recovery time invariably places it above the transition range. At the time of writing (2021) most lone Cubic and CReno flows have large enough recovery time to be above
the transition range, but a significant minority of CReno flows are within or even slightly
below this range. 

\subsection{Sawtooth Geometry}\label{pi2param_Sawtooth_Geometry}

\begin{figure*}
	\centering
	\includegraphics[width=\linewidth]{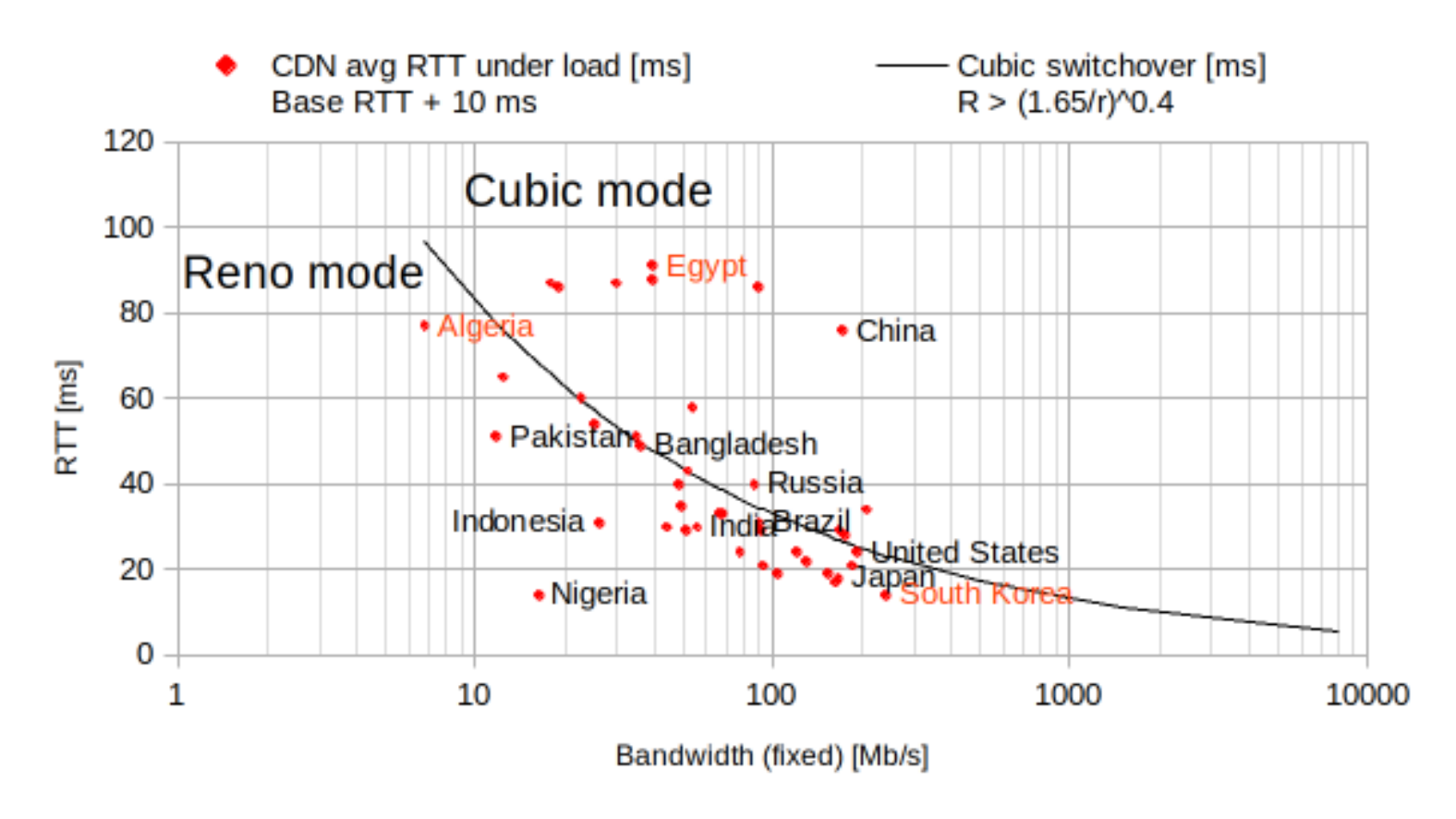}
	\caption{Scatter-plot per country of average user to CDN RTT under load against
	average fixed downstream access bandwidth. RTT is taken as if it is under load
	over the PI\(^2\) AQM under study, so RTT = base RTT plus 10\,ms (see text for explanation). Only the 43
	countries with the most Internet users are plotted, representing 90\% of
	Internet users. The top ten are labelled as well as those at the extremes. The
	curve overlaid on the plot is where the Cubic congestion control in Linux
	switches over from Reno mode to pure Cubic
	mode}\label{fig:pi2param-CDN-BDP-under-load}
\end{figure*}

The fraction, \(\lambda_0\), of the sawtooth amplitude that lies below the average
is important when determining the target queue delay. For an AIMD sawtooth,
\autoref{eqn:lambda} in \autoref{pi2param_reno_R_avg} gives a good approximation\footnote{See
	\autoref{pi2param_reno_R_avg} for the full approximation conditions.} as:
\[\lambda_0=\frac{(2+b_r)}{3(1+b_r)}.\] And, for \(b\ge\sfrac{1}{2}\) a sufficient
approximation is \(\lambda_0\approx\sfrac{1}{2}\). For a
Cubic sawtooth, \autoref{pi2param_cubic_R_avg} proves that 
\[\lambda_0=\sfrac{3}{4},\] whatever the value of \(b_c\).

The ``Great TCP Congestion Control Census'' \cite{Mishra19:CC_Census} conducted
by Mishra \emph{et al} in Jul--Oct 2019 found that Cubic was the most used by
nearly 31\% of the Alexa top 20k web sites, but BBR was approaching 18\%, and
already had a larger share of the Alexa top 250, as well as contributing 40\% by
downstream traffic share.\footnote{The census did not investigate congestion
controls used by QUIC.} Of the 51\% of the Alexa top 20k sites that were not
using either Cubic or BBR, 19\% were split between eight other known
controllers, the greatest shares being for YeAH and CTCP or Illinois at under
6\% each. The remaining 32\% were unidentifiable, including sites that were
unresponsive or did not serve anything large enough to be testable. As part of
that remaining 32\%, nearly 17\% of the total were using an unknown congestion
controller and further investigation found nearly 6\% of the total were using an
undocumented Akamai controller.

BBRv2~\cite{Cardwell17:BBR_ID} supports L4S when it detects ECN marking, so it
is unlikely to use the Classic queue. This leaves 67\% of sites that use some
form of Classic
congestion control, of which 46\% use Cubic and the remainder is split across a
dozen or so other algorithms, many of which, like Cubic, attempt to be friendly
to Reno at low BDP.

Based on recent predictions, more than two-thirds of Internet traffic now
emanates from Content Distribution Networks (CDNs) or cloud services distributed
to locations close to, and often within, the metro area or regional network of
the end-user's ISP~\cite{LaRoche20:CDNPerf}.

\autoref{fig:pi2param-CDN-BDP-under-load} illustrates how likely it is that Cubic congestion
control runs in its Reno mode for CDN traffic over a PI\(^2\)
AQM. The figure visualizes the average CDN RTT\footnote{Both fixed and mobile---the study did not
measure fixed and mobile separately. Nonetheless, RIPE Atlas probes are generally connected
to fixed access links although some are connected via Ethernet to mobile
broadband.} under load against average fixed downstream bandwidth per household.

To roughly model latency under load, the base RTT from
\autoref{pi2param_CDN_R_typ} is increased by \(\lambda_0*\mathtt{target}\) to
represent how much the average RTT would shift deeper into the queue until the
tips of the sawteeth would hit the PI\(^2\) AQM's \texttt{target}. Given the
Cubic congestion control switches into cubic mode as its BDP rises along the
sawtooth, there is a hybrid region where the bottoms of the sawteeth are CReno
and the tops Cubic. Therefore we set \(\lambda_0\) to the average of
\(\lambda_0\mathrm{creno}=9/17\) and \(\lambda_0\mathrm{cubic}=3/4\), that is
0.64. Hence the uplift of
\(0.64*15\,\mathrm{ms}\approx10\,\mathrm{ms}\).\footnote{It will not go
unnoticed that there is a circular dependency here, where we have to assume the
\texttt{target} chosen for the PI\(^2\) AQM as part of the process of
determining it.}

In order not to clutter the plot, the countries ranked highest by number of
Internet users are plotted until together they represent over 90\% of Internet
users. The countries labelled in black are the top 10 ranked by number of users
(see \autoref{pi2param_CDN_R_typ} for the detailed data and sources). As points
of interest, the countries at the extremes are also labelled, but in red.

It can be seen that a large proportion of Internet users sit 
at or below the upper limit of Cubic's Reno-friendly mode for a single flow. 
The countries above the switch-over curve other than China and Russia together
account for about 7\% of Internet users.
There is a question
mark over the CDN RTT in China, which might place China's point at lower RTT (see \autoref{pi2param_CDN_R_typ}). Given this unexplained but highly significant
outlier, for now we take the weighted average excluding China, which is 25\,ms.

As link rates continue to scale, the points are expected to shift inexorably to
the right. However, for some considerable time to come, Cubic could remain in
Reno mode for many users because, as CDN deployment continues and as focus
shifts to latency as well as bandwidth, the points are also expected to shift
downwards\footnote{This will continue to reduce the global typical RTT
(\texttt{RTT\_typ}) so that it will become possible to reduce the default
\texttt{target} of PI\(^2\), thus reducing the uplift of all the points, in turn
shifting more of them below the switch-over curve.} as they have already done in
the more mature deployments in N America, the Pacific rim and Europe (the larger
European countries are all in the cluster to the left of the US and Japan). Also
remember that we have chosen to examine the worst case of
a single downstream flow; whenever there are more simultaneous flows, the points
shift back to the left, into the Reno region. And, where capacity is
asymmetric, upstream flows also sit further to the left.

The curve overlaid on \autoref{fig:pi2param-CDN-BDP-under-load} shows the
relationship between throughput and RTT at the switch-over between CReno and
Cubic. It is derived from the formulae for the steady-state packet rate
in Cubic's Reno mode (\autoref{eqn:creno}) and in pure cubic
mode (\autoref{eqn:cubic}), as given below, assuming 1500B packets (\ref{itm:pi2param_one_long_flow}).

\begin{align}
r_\mathrm{creno} &= \frac{1}{R}\left({\frac{3}{2p}}\right)^{1/2}\label{eqn:creno}\\
r_\mathrm{cubic} &= \left(\frac{C(3+b_c)}{4(1-b_c)p^3R}\right)^{1/4}\label{eqn:cubic}
\intertext{At the same AQM loss probability, \(p\), the packet rate,
	\(r_\mathrm{cubic}\) equals \(r_\mathrm{creno}\) when the switchover RTT is}
R &= \left(\frac{27(1-b_c)}{2C(3+b_c)r^2}\right)^{1/5}
\end{align}
We can plug in the Cubic parameters recommended in
RFC8312~\cite{Rhee18:Cubic_RFC} and used in all known implementations, that is
multiplicative decrease factor \(b_c=0.7\); aggressiveness constant \(C=0.4\) in
cubic mode; and additive increase factor \(a_c=3(1-b_c)/(1+b_c)=0.53\) in
Reno-friendly mode.\footnote{Appx.\ A of \cite{DeSchepper16a:PI2} wrongly states
that Linux uses \(C=0.6\) and \(a_c = 1\), which leads to incorrect constants in
the resulting equations. \autoref{eqn:creno} \& \autoref{eqn:cubic_switch}
respectively correct equations (7) \& (8) in that paper.}\(^,\)\footnote{The
coefficient of 1.22 in \autoref{eqn:cubic_switch} is coincidentally the same to
2 dec. places as the \(\sqrt{3/2}\) coefficient in \autoref{eqn:creno}. However,
it can be seen from its derivation that it is unrelated.}
\begin{align}
  R &= \frac{1.22}{r^{2/5}}
     = \left(\frac{1.65}{r}\right)^{2/5}\label{eqn:cubic_switch}.
\end{align}

In summary, in the immediate future, the prevalent sawtooth geometry of Classic
traffic is likely to be dominated by the Reno mode of Cubic, with
\(\lambda_0\approx\sfrac{1}{2}\) and \(b_c=0.7\). 
Whether more traffic shifts to true Cubic geometry or stays as CReno depends on
whether, and how much, latency reduction overtakes bandwidth increase as the
predominant global trend in performance improvement.

\section{Typical Base RTT}\label{pi2param_RTT_typ}

A globally typical RTT for CDN traffic is calculated in
\autoref{pi2param_CDN_R_typ}. The average RTT for each country is weighted by
the Internet user population in that country, as collated on
Wikipedia~\cite{Wikipedia20:Inet_users_country} from multiple primary sources.
The countries are ranked in order of user population until 90\% of the total
Internet users in the world are covered. The CDN RTT per country is based on
measurements by Beganovi\'c in 2019 using RIPE Atlas probes deployed by self-selected
volunteers in what is claimed to be the largest Internet measurement
infrastructure in the world~\cite{Beganovic19:CDN_RTT}.

The resulting weighted average RTT to CDNs is 34\,ms. However, there is a
question-mark over some of the latency figures, given the measurements were all
taken to 7 CDNs with global coverage, which might not be representative of the
CDN market in certain countries). The data point for China is a particularly suspect outlier given the CDNs used for the measurements excluded all the top CDNs in China (see \autoref{pi2param_CDN_R_typ}). Therefore, we have decided to exclude it pending further investigation, given the weighted average is so sensitive to an error in this single data point. This results in a weighted average RTT to CDNs of 25ms.

As a sanity check, 25\,ms compares reasonably well with the global averages
given on Ookla's Speedtest Global Index page:
\begin{itemize}[nosep]
	\item 20\,ms fixed and 37\,ms mobile (Apr 2021 data);
	\item 24\,ms fixed and 42\,ms mobile (Apr 2020 data).
\end{itemize}
Ookla's data is collected from self-selecting users who use speedtest's
algorithm to find the closest CDN-based servers \cite{Ookla:Speedtest_index}.
The page gives a single global figure without details of the method used.

\section{Default \texttt{target}}\label{pi2param_target}

When selecting a global default for \texttt{target}, the aim is to ensure that
the AQM keeps queue delay reasonably low while not compromising utilization for
a large majority of users. If the \texttt{target} were set at the median delay,
it would cause under-utilization for half the global user population.
So ideally a latency figure for say the 75th or 90th
percentile of users would be used to derive \texttt{target}, but only data on
averages not percentiles is available globally (\S\,\ref{pi2param_RTT_typ}).

Therefore, a 'safety factor' is applied to the average RTT between users and
CDNs, which has to allow for the statistical distribution of RTTs to CDNs,
particularly for users in rural areas~\cite{Kulatunga15:Rural_Bloat}, who will
be further from the nearest CDN and who are also likely to have least bandwidth
and therefore be least willing to see it eaten by under-utilization. The safety
factor also has to allow for flows between clients and servers other than those in
CDNs. As an interim educated guess, we apply the safety factor, \(f=2\).

\balance
Next we draw together all the strands of the analysis of sawtooth scaling, positioning and
geometry in \S\,\ref{pi2param_Scaling}, in order to derive a default
\texttt{target}. To avoid underutilization for most users, we want \(f
R_\mathrm{typ}\) to sit at least at the minimum of the sawteeth,
\(R_\mathrm{min}\).

According to the discussion on sawtooth positioning in
\S\,\ref{pi2param_Sawtooth_Position}, BDPs have often, but not always, become
high enough that sawteeth will settle with their tips tending down towards the
target operating point of a PI\(^2\) AQM, rather than their average.

Therefore the fraction, \(\lambda\), of the amplitude that will be below
\texttt{target} can be used to relate \texttt{target} to the minimum RTT:
\begin{align*}
\mathtt{target}	&= \lambda (R_\mathrm{max} - R_\mathrm{min})\\
				&= \frac{\lambda(1-b)}{b} R_\mathrm{min},
\end{align*}
where \(\lambda\) has to be estimated from the actual geometric fraction of the
amplitude below the average, \(\lambda_0\), but also takes into account the
discussion in \S\,\ref{pi2param_Sawtooth_Position}. Therefore, finally we can
say:
\begin{equation}
	\boxed{\mathtt{target} \approx \frac{\lambda(1-b)f}{b} R_\mathrm{typ}}
\end{equation}

We call \(\sfrac{\lambda(1-b)}{b}\) the geometry factor. The geometry factors of
a selection of congestion controls (CCs) are tabulated below (Cubic in Reno mode
is abbreviated to CReno). Their geometry parameters are as recommended in the RFCs,
which all known implementations follow.
\begin{table}[h]
	\centering
	\begin{tabular}{rrrrrr}
	CC   & \(\lambda_0\)	& \(\lambda\) & \(b\)	&\(\sfrac{\lambda(1-b)}{b}\) & \(w\)\\
	\hline
	Reno & \(\sfrac{1}{2}\)	& 0.9 	& 0.5	& 0.90		&\\
	CReno& \(\sfrac{1}{2}\)	& 0.9 	& 0.7	& 0.39		& 70\%\\
	Cubic& \(\sfrac{3}{4}\)	& 0.85	& 0.7	& 0.36		& 30\%\\
	\hline
	\end{tabular}\label{tab:pi2param_geometry}
\end{table}

Taking account of the mix of congestion controls discussed in
\S\,\ref{pi2param_Sawtooth_Geometry}, but without modelling all the minor
players, we use a weighted average of CReno and Cubic using the weight, \(w\)
shown in the above table (based on the discussion in
\S\,\ref{pi2param_Sawtooth_Geometry}), which gives
a geometry factor of about 0.38. Thus, for PI\(^2\) we suggest setting the
default to:
\begin{align*}
\mathtt{target}	&= 0.38 * 2 * 25\,\mathrm{ms}\\
				&= 19\,\mathrm{ms}.
\end{align*}

Over time, as CDN deployment continues, \(R_\mathrm{typ}\) will continue to
reduce, evidenced by the latency reduction between 2020 and 2021 in the Ookla figures above: 17\% fixed, 12\% mobile. So the default
\texttt{target} could be reduced in future. That in turn will reduce RTT
further, with the knock-on effect of keeping more Cubic flows in Reno mode, thus
reinforcing the applicability of the lower \texttt{target} for AQMs.

Other implementations intended for particular link technologies might use a
different default today. For instance, the Low Latency DOCSIS
specification~\cite{CableLabs:DOCSIS3.1} uses
\(\mathtt{target}=10\,\mathrm{ms}\), which perhaps makes sense because cable
technology is less likely to extend to rural areas, so the distribution around
the average RTT is likely to be considerably tighter. By a similar argument, the
default \texttt{target} for mobile networks might need to be greater than
recommended here, depending on how well 5G meets its aspirations to reduce base RTT.

Of course, operators are free not to use the default \texttt{target} for
out-of-the-ordinary environments. For instance, they could configure a higher
\texttt{target} for satellite links and remote rural locations; or a lower
\texttt{target} for highly concentrated urban deployments. Nonetheless, the
purpose of this report has been to recommend a default that would be suitable
across the Internet.
\section{Acknowledgements}\label{pi2param_acks}

Thanks to Vidhi Goel for pointing out the need to use RTT under load in the
Cubic switch-over scatter-plot; to Asad Sajjad Ahmed for the empirical plots; to
Neal Cardwell for pointing out the erroneous parameters used for Linux Cubic;
and to Koen De Schepper for pointing out the need to consider sawtooth geometry
and for pointing out the significance of the max RTT of the AQM, not just
\texttt{Tupdate}, in the sawtooth position analysis. The author alone is to
blame for any remaining errors.

\clearpage
\nobalance
\appendix
\section{Average Queue Over a Reno Sawtooth}\label{pi2param_reno_R_avg}

The following analysis determines the fraction \(\lambda_0\) of the amplitude of a
Reno sawtooth that sits below its average. It is generalized for any additive
increase of \(a\) segments per round (which may be fractional). Terminology and assumptions are defined in
the body of the paper (\S\,\ref{pi2param_terms} \& \S\,\ref{pi2param_Scaling}).

Reno's congestion window increases by \(a\) segments per round,
\begin{align*}
W_r(j) = W_\mathrm{min} + ja,
\end{align*}
where \(j\) is an index of the rounds since the last reduction. By the same
reasoning as in \S\,\ref{pi2param_Scaling}, while the link is not underutilized,
Reno's RTT is directly proportional to its congestion window:
\begin{align*}
R_r(j) = R_\mathrm{min} + \frac{ja}{r},
\end{align*}
where \(r\) is the packet rate, so \(a/r\) is the delay added to the queue by one round of additive increase.
For brevity we will use \(R_a = a/r\) to denote this addition to the RTT per
round. By our assumption that the queue is never allowed to drain completely, we
can remove the minimum queue delay from the equation and focus solely on the
additional delay above the minimum,
\begin{align*}
d(j) = jR_a.
\end{align*}
Then the fraction of the amplitude that sits below the average,
\begin{align*}
\lambda_0 &= \frac{\mathbb{E}\{d(t)\}}{d_\mathrm{max}}\\
        &= \frac{\mathbb{E}\{d(t)\}}{R_\mathrm{min}(1-b)/b}.
\end{align*}
Within a cycle, the queue above the minimum averaged over time,
\(\mathbb{E}\{d_r(t)\}\), is the extra queue above the minimum in each round
weighted by the duration of each round then divided by the sum of the weights.
The duration of each round is the RTT itself. Thus,
\begin{align*}
\lambda_0 &= \frac{\left.\sum_{j=0}^{J-1}(R_\mathrm{min} + jR_a)jR_a\middle/\sum_{j=0}^{J-1}(R_\mathrm{min} + jR_a)\right.}{R_\mathrm{min}(1-b)/b}
\intertext{Approximating \(J(J-1)\) as \(J^2\), and using the standard result for a sum of squares without approximation\bob{ToDo: Mixed approximations}, then simplifying:}
        &\approx \frac{J^2 R_\mathrm{min} R_a/2 + (J^3/3+J^2/2+J/6)R_a^2}{(JR_\mathrm{min}+J^2R_a/2)R_\mathrm{min}(1-b)/b}\\
        &\approx \frac{3J R_\mathrm{min} R_a + (2J^2+3J+1)R_a^2}{(6R_\mathrm{min}^2+3JR_\mathrm{min}R_a)(1-b)/b}
\end{align*}
The maximum value of \(j\) under stable conditions can be found by equating the
additive increase over a cycle to the multiplicative decrease,
\begin{align*}
J R_a &= R_\mathrm{min} (1-b)/b.
\end{align*}
Substituting for \(J\), then collecting terms and simplifying further,
\begin{align}
\lambda_0 &= \frac{3R_\mathrm{min}^2 + (2R_\mathrm{min}^2(1-b)/b+3R_\mathrm{min}R_a+R_a^2b/(1-b)}{(6R_\mathrm{min}^2+3R_\mathrm{min}^2(1-b)/b)}\notag\\
        &= \frac{(2+b)/b + 3R_a/R_\mathrm{min} + b/(1-b) R_a^2/R_\mathrm{min}^2}{3(1+b)/b}\notag\\
        &= \frac{(2+b)}{3(1+b)} + \frac{b}{(1+b)}\frac{R_a}{R_\mathrm{min}} + \frac{b^2}{3(1-b^2)}\left(\frac{R_a}{R_\mathrm{min}}\right)^2\notag\\
        & \approx \frac{(2+b)}{3(1+b)}\qquad \mathrm{if}\quad R_a\ll \left.R_\mathrm{min}(1+b)\middle/b\right.\label{eqn:lambda}.
\end{align}
Then, for standard Reno with \(b=\sfrac{1}{2}\),
\[\lambda_0\approx\sfrac{5}{9}\approx0.556.\] And for Cubic-Reno with \(b=0.7\),
\[\lambda_0\approx\sfrac{9}{17}\approx0.529.\] \(\lambda_0\approx\sfrac{1}{2}\) is a
good enough approximation for many purposes, including for the present paper.

\section{Average Queue Over a Cubic Sawtooth}\label{pi2param_cubic_R_avg}

The following analysis determines the fraction \(\lambda_0\) of the amplitude of a
Cubic sawtooth that sits below its average. Terminology and assumptions are
defined in the body of the paper (\S\,\ref{pi2param_terms} \&
\S\,\ref{pi2param_Scaling}).

The formula for the congestion window of a Cubic sawtooth is defined in IETF RFC
8312~\cite{Rhee18:Cubic_RFC} as,
\begin{align*}
W_c(t) = W_\mathrm{cmax} + C(t-K)^3,
\end{align*}
where \(C\) is a constant (0.4 in known implementations, as recommended in RFC 8312~\cite{Rhee18:Cubic_RFC}) and:
\begin{align*}
K = \left(\frac{W_\textrm{cmax}(1-b)}{C}\right)^\frac{1}{3},
\end{align*}
where \(b\) is the multiplicative decrease factor already defined in
\S\,\ref{pi2param_terms} (recommended and implemented as 0.7).

By the same reasoning as in \S\,\ref{pi2param_Scaling}, while the link is not
underutilized, Cubic's RTT is directly proportional to its congestion window:
\begin{align*}
R_c(t) = R_\mathrm{cmax} + \frac{C(t-K)^3}{r},\\
K = \left(\frac{rR_\textrm{cmax}(1-b)}{C}\right)^\frac{1}{3}.
\end{align*}

The average RTT over a cycle, \(\mathbb{E}\{R_c(t)\}\), is then\footnote{A continuous integral rather than discrete sum is a sufficient approximation.}
\begin{align}
R_{c0} &= \frac{1}{K} \int^K_0 R_\mathrm{cmax} + \frac{C(t-K)^3}{r} dt,\notag\\
&= \frac{1}{K} \left[R_\mathrm{cmax}t + \frac{C(t-K)^4}{4r}\right]^K_0\notag\\
&= R_\mathrm{cmax} - \frac{CK^3}{4r}\notag\\
\intertext{Substituting for \(K\):}
&= R_\mathrm{cmax} \frac{(3+b)}{4}.\label{eqn:R_avg_cubic}
\end{align}

Then, for a single Cubic sawtooth, the fraction of the amplitude that sits below
the average is
\begin{align}
\lambda_0c &= \frac{(R_\mathrm{c0}-R_\mathrm{cmin})}{(R_\mathrm{cmax} - R_\mathrm{cmin})}\notag\\
&= \frac{R_\mathrm{cmax} \left(\frac{(3+b)}{4}-b\right)}{R_\mathrm{cmax}(1-b)}\notag\\
&= \frac{3}{4}\label{eqn:lambda_cubic}.
\end{align}

Thus, \(\lambda_0c\) is constant for any \(b\in [0,1)\).

\section{Scaling of AIMD Cycle Duration}\label{pi2param_Scaling_Cycle}

Scaling of the cycle duration with flow rate is not directly relevant to the setting of
PI\(^2\) parameters, but it does affect utilization in two indirect but important
ways: 
\begin{itemize}[nosep]
\item As flow rate scales, cycle duration increases relative to the fixed update
time of the PI\(^2\) AQM. over a transition range of flow rates, the queue delay
sawtooth shifts down relative to the AQM target (see
\S\,\ref{pi2param_Sawtooth_Position}), potentially leading to poorer utilization
at rates above the transition.
\item A Classic congestion control responds to a single loss or ECN mark, so
losses and ECN marks have to be completely absent during a cycle for a flow to
maintain full utilization. The longer the duration of each cycle, the more
likely that some extraneous event will occur, e.g.\ the arrival of a brief flow
or loss due to a transmission error. This noise sensitivity of Classic flows
becomes the dominant determinant of utilization the more flow rate scales (see
footnote 6 of Jacobson \& Karels~\cite{Jacobson88b:Cong_avoid_ctrl}).
\end{itemize}
	
Under the same assumptions as defined in \S\,\ref{pi2param_Scaling}, cycle
duration (or recovery time), \(T_r\), depends on base RTT, \(R_b\), and packet
rate, \(r\), as follows.

For an AIMD congestion control the increase and decrease balance in steady state, 
\begin{align}
	Ja	&= W_\mathrm{max} - W_\mathrm{min}\notag\\
		&= W_\mathrm{min} (1/b - 1)\notag\\
		&= rR_\mathrm{min} \frac{(1 - b)}{b}\notag\\
	J	&= rR_\mathrm{min} \frac{(1 - b)}{ab}\label{eqn:Jaimd}
\end{align}
where \(J\) is the number of rounds between reductions. Then the recovery time,
\begin{align}
	T	&= \sum_{j=0}^{J-1} \left(R_\mathrm{min} + \frac{ja}{r}\right)\notag\\
		&= J R_\mathrm{min} + \frac{J^2a}{2r},\notag
\intertext{approximating \(J(J-1)\) as \(J^2\), then substituting for \(J\) from \autoref{eqn:Jaimd},}
		&= rR_\mathrm{min}^2 \left(\frac{(1 - b)}{ab} + \frac{(1-b)^2}{2ab^2}\right)\notag\\
		&= rR_\mathrm{min}^2 \frac{(1-b^2)}{2ab^2}\label{eqn:TaidmRmin}.
\intertext{And from \autoref{eqn:Rmin_max} \& \autoref{eqn:R0_max}}
		&= rR_\mathrm{0}^2 \frac{(1-b^2)}{2a(\lambda_0+b-\lambda_0 b)^2}\notag
\intertext{and from \autoref{eqn:lambda},}
		&= rR_\mathrm{0}^2 \frac{9(1+b)^3(1-b)}{8a(1+b+b^2)^2}.
\intertext{For Reno, \(a_r=1, b_r=1/2\):}
	T_r	&= \frac{3}{2}rR_\mathrm{min}^2\notag\\
		&= \frac{3}{8}rR_\mathrm{max}^2\notag\\
		&= \frac{243}{392}rR_\mathrm{0}^2 
			\qquad\qquad\qquad\approx 0.62rR_\mathrm{0}^2\label{eqn:Treno}.
\intertext{For CReno, \(a_c=3(1-b_c)/(1+b_c), b_c=0.7\):}
	T_c	&= \frac{(1+b_c)^2}{6b^2}rR_\mathrm{min}^2\notag\\
		&= \frac{289}{294}rR_\mathrm{min}^2
			\quad\qquad\qquad\approx 0.98\,rR_\mathrm{min}^2\notag\\
		&= \frac{289}{600}rR_\mathrm{max}^2
			\quad\qquad\qquad\approx 0.48\,rR_\mathrm{max}^2\notag\\
		&= \frac{3(1+b_c)^4}{8(1+b+b^2)^2}rR_\mathrm{0}^2
			\quad\enspace\approx 0.65\,rR_\mathrm{0}^2\label{eqn:Tcreno}.
\end{align}

As a double check, the recovery times in terms of \(R_0\) should be roughly the
same for Reno and CReno, by design (that for CReno should be a little greater,
because it is less curved).

This scaling of cycle duration is important to understand, as follows:
\begin{itemize}[nosep]
	\item Additive increase of a constant amount of data per round trip causes the
	duration of a single flow's sawtooth cycle to double for every doubling of link
	rate. This can be seen for a) Reno and b) Cubic in Reno mode in the right-hand
	column of \autoref{fig:pi2param-scaling}. But for every doubling of the RTT
(whether min, max or mean), the duration of each cycle quadruples, as
illustrated by \autoref{eqn:TaidmRmin} or its subsequent variants. This is
because it takes double the number of RTTs to regain its window, but also each
RTT is double the length.
	
	\item In contrast, the cycle duration of a purely Cubic congestion control
	scales with the cube-root of bandwidth-delay product (BDP) \cite{Rhee18:Cubic_RFC}. So, as link
	capacity or RTT doubles, the duration of the cycles of a single flow grows by
	\(2^{1/3} \approx 1.26\), as can also be seen for c) Cubic in the right-hand
	column of \autoref{fig:pi2param-scaling}.
\end{itemize}

Note, though, that the \emph{amplitude} of Cubic's queue-delay variation still
scales like Reno, i.e.\ linearly with RTT and invariant with link capacity,
because it is determined by the multiplicative decrease.

\section{Typical User to CDN RTT}\label{pi2param_CDN_R_typ}

\balance
Beganovi\'c~\cite{Beganovic19:CDN_RTT} provides the average RTT measured using
ICMP ping from probes in each country to sites known to be served by CDNs. The
data was collected from RIPE Atlas probes deployed by volunteers around the
world, and was last updated on 17 Apr 2019.%

\begin{figure*}
	\centering
	\includegraphics[width=\linewidth]{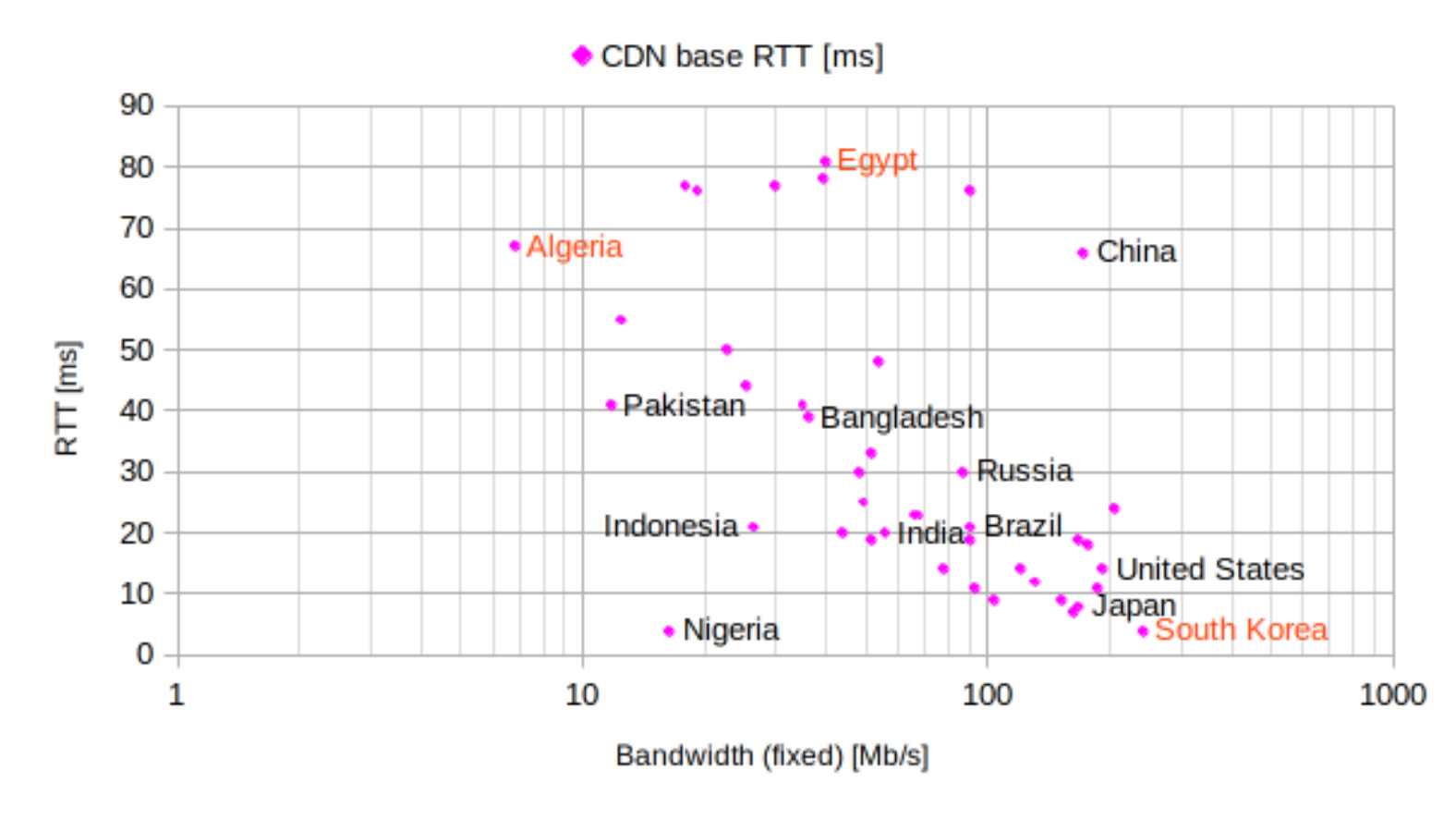}
	\caption{Scatter-plot per country of average base RTT from users to CDNs and
		average fixed access bandwidth. Only the 43 countries with the most Internet
		users are plotted, representing 90\% of Internet users. The top 10 are labelled
		as well as those at the extremes}\label{fig:pi2param-CDN-BDP}
\end{figure*}

The data is tabulated below and visualized in \autoref{fig:pi2param-CDN-BDP}. At
the bottom of the table, an average is derived, weighted by the population of
Internet users in each country (taking the countries with the highest Internet
user populations until 90\% of the world's total Internet users are covered).
The per-country data on numbers of Internet users was taken from
Wikipedia~\cite{Wikipedia20:Inet_users_country}, which in turn used population
figures for each country, usually from the US Census Bureau, and various
estimates of the percentage of Internet users in each country, mostly provided
by the ITU.

The measurements were taken to the following seven global CDNs:
\begin{itemize}[nosep]
	\item Akamai
	\item AWS Cloudfront
	\item Microsoft Azure
	\item Cloudflare
	\item Google Cloud CDN
	\item Fastly
	\item Cachefly
\end{itemize}

The data point for China seems uncharacteristic for countries of similar size
and market maturity. It is possible that it is suspect, perhaps because
measurements to large Chinese CDN providers such as the following were not
included in the RIPE Atlas study: `
\begin{itemize}[nosep]
	\item Alibaba Cloud
	\item Baidu Cloud
	\item BaishanCloud
	\item ChinaCache
	\item Tencent Cloud
\end{itemize}
Given users in China make up nearly a quarter of the global total, the weighted
average would be sensitive to any large error in the CDN latency for users in
China. For instance, if the latency figure just for China was reduced from 66ms
to 20ms (bringing it in line with India), the global weighted average would drop
from 34ms to 24ms. Therefore, for now we exclude the data point for China,
resulting in a weighted average CDN latency of 25\,ms.

\autoref{fig:pi2param-CDN-BDP} shows the countries ranked highest by number of
Internet users until together they contain over 90\% of Internet users. This
avoids cluttering the plot. Further, the countries labelled in black are the top
10 ranked by number of users.  The stripe of points with higher RTT than 66\,ms 
together represent less than 5\% of the total users in the plot. If the EU were
one country its point would sit in the cluster to the left of the US and Japan,
which represent the larger countries in Europe. As points of interest, the
countries at the extremes are labelled in red (Nigeria is both in the top 10 by 
user population, and it has the lowest latency, tying with South Korea).

\begin{table*}
	\small
	\begin{tabular}{p{2.5cm}rrrcc}
		Country&Population &\% of&Internet users&Fixed bandwidth (Mb/s)&CDN latency (ms)\\
&&popul'n&\cite{Wikipedia20:Inet_users_country}&\cite{Ookla:Speedtest_index}&\cite{Beganovic19:CDN_RTT}\\
\hline
China&1,427,647,786&69.27\%&988,990,000&172.95&66\\
India&1,366,417,754&55.31\%&755,820,000&55.76&20\\
United States&324,459,463&96.26\%&312,320,000&191.97&14\\
Indonesia&266,911,900&79.56\%&212,354,070&26.31&21\\
Brazil&213,300,278&75.02\%&160,010,801&90.3&21\\
Nigeria&205,886,311&66.15\%&136,203,231&16.33&4\\
Russia&143,989,754&82.39\%&118,630,000&87.01&30\\
Japan&127,484,450&91.27\%&116,350,000&167.18&8\\
Bangladesh&164,945,471&70.41\%&116,140,000&36.02&39\\
Pakistan&213,756,286&47.10\%&100,679,752&11.74&41\\
Mexico&128,972,439&69.01\%&89,000,000&48.35&30\\
Iran&83,020,323&94.06\%&78,086,663&19.17&76\\
Germany&82,114,224&94.74\%&77,794,405&120.93&14\\
Philippines&104,918,090&69.58\%&73,003,313&49.31&25\\
Vietnam&97,338,579&70.04\%&68,172,134&66.38&23\\
United Kingdom&66,181,585&98.22\%&65,001,016&92.63&11\\
Turkey&80,745,020&76.88\%&62,075,879&34.95&41\\
France&64,979,548&89.32\%&58,038,536&192.25&14\\
Egypt&101,545,209&53.91\%&54,740,141&39.66&81\\
Italy&60,416,000&83.65\%&50,540,000&90.93&19\\
South Korea&50,982,212&96.94\%&49,421,084&241.58&4\\
Spain&46,750,321&90.70\%&42,400,756&186.4&11\\
Thailand&69,037,513&52.89\%&36,513,941&206.81&24\\
Poland&38,382,576&90.40\%&34,697,848&130.98&12\\
Canada&36,624,199&92.70\%&33,950,632&167.61&19\\
Argentina&44,271,041&75.81\%&33,561,876&51.51&19\\
South Africa&56,717,156&56.17\%&31,858,027&43.91&20\\
Colombia&49,065,615&62.26\%&30,548,252&53.73&48\\
Ukraine&44,222,947&66.64\%&29,470,000&67.52&23\\
Saudi Arabia&32,938,213&82.12\%&27,048,861&90.24&76\\
Malaysia&31,624,264&80.14\%&25,343,685&103.34&9\\
Morocco&35,739,580&61.76\%&22,072,765&25.37&44\\
Taiwan&23,626,456&92.78\%&21,920,626&163.85&7\\
Australia&24,450,561&86.54\%&21,159,515&77.88&14\\
Venezuela&31,977,065&64.31\%&20,564,451&17.9&77\\
Algeria&41,318,142&47.69\%&19,704,622&6.78&67\\
Ethiopia&104,957,438&18.62\%&19,543,075&12.39&55\\
Iraq&38,274,618&49.36\%&18,892,351&29.88&77\\
Uzbekistan&31,910,641&52.31\%&16,692,456&39.2&78\\
Myanmar&53,370,609&30.68\%&16,374,103&22.75&50\\
Netherlands&17,035,938&93.20\%&15,877,494&152.94&9\\
Peru&32,165,485&48.73\%&15,674,241&51.81&33\\
Chile&18,054,726&82.33\%&14,864,456&176.48&18\\
\hline
&&\% world&&\multicolumn{2}{c}{Averages weighted by Internet users}\\
Above countries&&90.15\%&4,292,105,058&103.32&34\\
\multicolumn{2}{l}{Above countries excl.\ China}&69.37\%&3,303,115,058&82.47&25\\
\hline
World&&100.00\%&4,761,334,541&&\\

	\end{tabular}
\end{table*}

` 
\balance%
\addcontentsline{toc}{section}{References}

\newpage
{%
\scriptsize%
\bibliography{pi2}}

\onecolumn%
\section*{Document history}

\begin{tabular}{|c|c|c|p{3.5in}|}
 \hline
Version &Date &Author &Details of change \\
 \hline\hline
00A &01-Jun-2021&Bob Briscoe &First draft\\\hline%
01 &02-Jun-2021&Bob Briscoe &Changed \autoref{fig:pi2param-CDN-BDP-under-load} to RTT under load. Numerous minor corrections.\\\hline%
01A &04 Jun 2021    &Bob Briscoe &Minor corrections following review.\\\hline%
02			&05 Jul 2021	&Bob Briscoe &Following review by Koen De Schepper, altered terminology, clarified different RTTs, added more rationale and added avg Reno qDelay appendix.\\\hline%
02A 		&19 Oct 2021    &Bob Briscoe &Fixed misconception: Linux CReno AI factor is 9/17, not 1 and Cubic aggressiveness is C=0.4, not 0.6. Added new section on sawtooth positioning and reworked rest of paper accordingly.\\\hline%
02B &21 Oct 2021    &Bob Briscoe &Fixed minor errors.\\\hline%
\metaversion &\metadate    &Bob Briscoe &Updated sawtooth positioning analysis, justified approximations in appendices, added small selection of empirical plots, added acks.\\\hline%
\hline%
\end{tabular}

\end{document}